\documentstyle[aps,prl,epsfig,pifont]{revtex}
\begin{document}
\twocolumn[\hsize\textwidth\columnwidth\hsize \csname
 @twocolumnfalse\endcsname

\title{Fluctuation-driven insulator-to-metal transition in an external
  magnetic field}

\author{V.  Jani\v{s}\ $^{a}$ and G. Czycholl\ $^{b}$}

\address{$^{a}$Institute of Physics, Academy of Sciences of the Czech
  Republic, Na Slovance 2, CZ-18221 Praha 8, Czech Republic\\
  $^{b}$Institut f\"ur Theoretische Physik, Universit\"at Bremen, PO Box
  330440, D-28334 Bremen, Germany}

\date{\today}
\maketitle 
\begin{abstract}
  We consider a model for a metal-insulator transition of correlated
  electrons in an external magnetic field. We find a broad region in
  interaction and magnetic field where metallic and insulating (fully
  magnetized) solutions coexist and the system undergoes a first-order
  metal-insulator transition. A global instability of the magnetically
  saturated solution precedes the local ones and is caused by collective
  fluctuations due to poles in electron-hole vertex functions.
  
\end{abstract}

] 
Metal-insulator together with magnetic transitions belong to the most
important collective phenomena caused by the (screened) Coulomb electron
repulsion.  Particularly the correlation-driven Mott-Hubbard transition
from a paramagnetic metal to a paramagnetic insulator, i.~e. with no
apparent symmetry breaking, has been studied intensively in an effort to
explain qualitatively the experimental findings in vanadium oxides doped
with chromium \cite{Gebhard97}. Although observed only at finite
temperatures, the Mott-Hubbard transition is theoretically studied mostly
as an extrapolation to zero temperature and in the mean-field limit with
only local dynamical fluctuations.  Antiferromagnetic long-range order is
supposed to be suppressed at low temperatures by a frustration destroying
the perfect-nesting property \cite{Georges96}.

But even in this maximally simplified situation one is unable to solve the
problem exactly. A scenario for the Mott-Hubbard transition in the
spin-symmetric case was initially deduced from the so-called iterated
perturbation theory (IPT) \cite{Zhang93}, but it was later supported by
analytical \cite{Moeller95} and numerical \cite{Bulla99}
renormalization-group arguments.  The detailed transition behavior remains,
however, still controversial \cite{Kehrein98}.

Unfortunately we do not have many cogent methods for the description of the
critical behavior at the metal-insulator transition. The existing solutions
are extrapolations. There is not a reliable interpolation between the
Fermi-liquid regime and the strong-coupling, atomic like solution.  IPT
seems the only technique based on perturbation theory being able to cover
the desired features of the weak and strong coupling limits and to describe
 qualitatively the Mott-Hubbard transition.

The description of the Mott-Hubbard transition in the spin-symmetric phase
is further plagued with the nonexistence of a parameter controlling the
critical behavior. Nozi\`eres \cite{Nozieres98} recently suggested to
extend the $d=\infty$ description of the Mott-Hubbard transition to the
magnetic case.  The density of states (DOS) is no longer pinned at the
Fermi energy and IPT does not reproduce correctly the strong-coupling limit
\cite{note0}. New methods are to be used to examine the magnetic case.

On the other hand, magnetic field and spin polarization offer a natural
parameter, magnetization, with the aid of which we can control the
transition behavior. Moreover, the exact insulating fully spin-polarized
state is explicitly known, namely the Hartree solution with well separated
bands for up and down spins. The question whether the saturated magnet
decays to a metallic state or whether the insulating phase remains stable
for non-saturated magnetizations at intermediate and strong coupling can be
studied as an extension of the Mott metal-insulator transition.

A picture for the Mott transition near the fully saturated magnetic state
in finite dimensions was suggested in \cite{vanDongen94}. It was shown that
the fully saturated ferromagnet decays at $B=B_s(U)$ differently at weak
and strong coupling. While in the former it is noncritical, a two-particle
function gets critical at the ferromagnetic boundary in the latter.
Recently this analysis was extended to the mean-field limit of a simplified
Hubbard model with one spin species static \cite{vanDongen97}.

Unlike the spin-symmetric case, it is the metallic side of the transition
that is difficult to describe in the magnetic case. There is no complete
solution below the ferromagnetic boundary at intermediate and strong
coupling. Only recently it was shown that for $B \nearrow B_s$ the parquet
approximation reproduces the exact asymptotics of the Bethe ansatz solution
in $d=1$ \cite{Janis98a}. However, a closed solution beyond weak-coupling
\cite{Jelitto71,Li93} has not yet been constructed.

The aim of this work is to investigate the existence and kind of the
fluctuation-driven insulator-to-metal transition in an applied magnetic
field within the exact limit of high spatial dimensions. We concentrate on
nearly ferromagnetic states at intermediate coupling.

Using field-theoretic, diagrammatic techniques we show that at weak
coupling, $U<U_{c_0}\approx 1.38w$ (for a semi-elliptic density of states
with the bandwidth $2w$), the insulating, fully magnetized solution goes
continuously over to a metallic one by merging the bands of the up- and
down-spin electrons at $B_s(U)=w-U/2$. The behavior is controlled by
one-particle functions.  At intermediate and strong coupling, $U>U_{c_0}$,
however, the magnetically saturated solution gets \textit{globally}
unstable along $B_0(U)>w-U/2$.  A second critical point
$U_{c_1}\approx1.65$ is found from which the saturated magnet gets also
locally unstable with a diverging electron-hole vertex function at
$B_s>w-U/2$ but $B_s<B_0$.

The metallic solution at intermediate coupling is stabilized by
correlations and the spectral function shows two satellite bands and a
Kondo-like narrow quasiparticle peak around the Fermi energy.  The metal
remains stable also for $B>B_s(U)$ where already the fully spin
polarized solution exists. It is destroyed at an upper critical filed
$B_u(U)$ with a jump in magnetization to the fully magnetically saturated
state.  We hence find a rather broad area of parameters $U,B$ for which the
metallic and insulating solutions coexist as local minima of the
free-energy functional and within which a \textit{first-order
  metal-insulator} transition takes place. We have a situation with
magnetic hysteresis and discontinuities in magnetization as known from
metamagnetic transitions \cite{Yamada93,Vollhardt97}. 

Unlike weak coupling, the insulator-to-metal transition at intermediate and
strong coupling is controlled by \textit{two-particle} vertex functions and
their singularities (poles). No weak-coupling theory suppressing vertex
corrections is able to describe this fluctuation-driven metal-insulator
transition and the metallic solution with a narrow quasiparticle resonance
correctly. Only self-consistent theories containing at least
\textit{infinite} series of multiple singlet and triplet electron-hole
scatterings can offer an appropriate qualitative picture of the behavior of
correlated electrons near the metal-insulator transition at intermediate
coupling.

We start with the Hubbard Hamiltonian in an external magnetic field $B$
\begin{eqnarray}
  \label{eq:hh}
  \widehat{H}&=&\sum_{{\bf k}\sigma} \left(\epsilon({\bf k})
    -\mu+\sigma B\right) c^{\dagger}_{{\bf k}\sigma}
  c^{\phantom{\dagger}}_{{\bf k}\sigma}  +
  U\sum_{{\bf i}}\widehat{n}_{{\bf i}\uparrow}\widehat{n}_{{\bf i}
    \downarrow},
\end{eqnarray}
with $\sigma=\pm1$ denoting the spin direction and $\mu$ the chemical
potential.  We are interested in half filling at which $\mu=U/2$ due to the
electron-hole symmetry. The one-particle propagator can be represented
as
\begin{eqnarray}
  \label{eq:GF-Hubbard}
   G_\sigma({\bf k},z)&=&\left[z+\sigma \left(B+\frac U2m\right)-\epsilon({\bf
       k}) -\Sigma_\sigma({\bf k},z)\right]^{-1} 
\end{eqnarray}
where $\Sigma_\sigma$ is a dynamical correction to the Hartree static
self-energy absorbed in the magnetization $m$. The dynamical self-energy is
determined from the Schwinger-Dyson equation of motion
\begin{eqnarray}
  \label{eq:sigma-2P}
  &&\Sigma_\sigma(k)=\\ 
  &&-\frac{U}{\beta^2N^2}\sum_{k'q}\Gamma_{\sigma-\sigma}(k,k';q)
  G_\sigma(k+q)G_{-\sigma}(k'+q)G_{-\sigma}(k').\nonumber 
\end{eqnarray}
where $\Gamma_{\sigma-\sigma}$ is the full two-particle vertex. We use a
four-vector notation $k=({\bf k},i\omega_n)$, $q=({\bf q},i\nu_m)$ for
fermionic and bosonic variables, respectively.

The vertex function $\Gamma_{\sigma-\sigma}$ has to be determined from a
diagrammatic skeleton expansion. To allow for multiple electron-hole
scatterings we must take into considerations at least the FLEX-type
approximations with singlet and triplet electron-hole channels
\cite{Janis98a,Bickers89}.  The electron-hole vertices are for the
singlet and triplet channels, respectively:
\begin{mathletters}
\begin{eqnarray}
  \label{eq:Gamma-eh}
  \Gamma_{\sigma-\sigma}(k,k';q)&=&\frac U{1+X_{\sigma-\sigma}(k-k')}ß , \\
  \label{eq:cal_K-eh}
  {\cal K}_{\sigma-\sigma}(k,k';q)&=&\frac U{1-X_{\uparrow\uparrow}(q)
    X_{\downarrow\downarrow}(q)} 
\end{eqnarray}
with dimensionless electron-hole bubbles
\begin{eqnarray}
  \label{eq:2P-bubble}
  X_{\sigma\sigma'}(q)&=&\frac U{\beta N}\sum_{k}
  G_\sigma(k) G_{\sigma'}(k+q)\ . 
\end{eqnarray}
\end{mathletters}
In this and the other FLEX approximations, the vertex at the two-particle
scattering events remains unrenormalized. These approximations are hence
reliable only at weak or moderate interaction strengths. At strong coupling
we have to pass to a higher level, the parquet approximation with
renormalized vertices at scattering events. At intermediate coupling, for
$U\approx 2w$, the FLEX and the two-channel parquet approximations deliver
qualitatively similar pictures in the spin-symmetric situation
\cite{Janis99a}.

The electron-hole bubble functions from (\ref{eq:2P-bubble}),
$X_{\sigma\sigma'}(q)$, are real and negative at the Fermi energy and the
vertex functions get singular and negative if
\begin{eqnarray}
  \label{eq:instability-nonlocal}
  1+X_{\sigma\sigma'}({\bf q},0)&\le& 0 \ .
\end{eqnarray}
Equality indicates the existence of a pole at the Fermi energy and
negativity an instability of the solution, i.~e. a local energy minimum
turns into a local maximum.

Bipartite lattices with perfect nesting and one electron per lattice site
get usually first unstable in the singlet channel at the antiferromagnetic
point ${\bf q}=(\pi,\pi,\ldots)$. Frustration due to the inclusion of
next-to-nearest or random hopping destroys this instability and a fully
frustrated model is expected to have no preference for a particular
low-temperature long-range order \cite{Georges96}.  In the next steps we
use this idealization of a fully frustrated system within the dynamical
mean-field theory where the four-momenta are substituted by Matsubara
frequencies.  Further on, we use a semi-elliptic model density of states of
the $d=\infty$ Bethe lattice $\rho_0(E)=\frac2{\pi w^2}\sqrt{w^2-E^2}$,
with $w=1$, approximating a three-dimensional DOS with a finite bandwidth
and sharp edges.

As a first step we investigate a local instability of the Hartree solution
that can occur only in the singlet channel, since the triplet bubbles
vanish in the saturated state.  It is easy to find the explicit
representation for the singlet electron-hole bubble
\begin{mathletters}\label{instability}
  \begin{eqnarray}
    \label{eq:HF-MF-gamma}
   X_{\uparrow\downarrow}(0)=-4U\left\{\frac 2{3\pi}\left[1
        -B_m^2\right]^{3/2}+B_m m\right\}&&\nonumber\\[2pt]
   +4U\int_{-1}^{-1+2B_m}\frac{dx}\pi\sqrt{1-x^2}\sqrt{(2B_m-x)^2-1}&&    
  \end{eqnarray}
  with the equilibrium magnetization
  \begin{eqnarray}
    \label{eq:HF-MF-m}
   m&=&\frac 2\pi\left[\arcsin\left(B_m\right)+B_m\sqrt{1-B_m^2}\ \right] .  
  \end{eqnarray}
\end{mathletters}
Here $B_m=B+\frac U2m$. The saturated magnetic solution ($m=1$) gets
unstable if either $B+U/2<1$ or $X_{\uparrow\downarrow}(0)<-1$. The
former condition follows from (\ref{eq:HF-MF-m}), i.~e. from the
one-particle magnetization. The latter is the existence criterion
for a pole in the vertex function $\Gamma_{\uparrow\downarrow}$.

\begin{figure}\hspace*{-10pt}
  \epsfig{figure=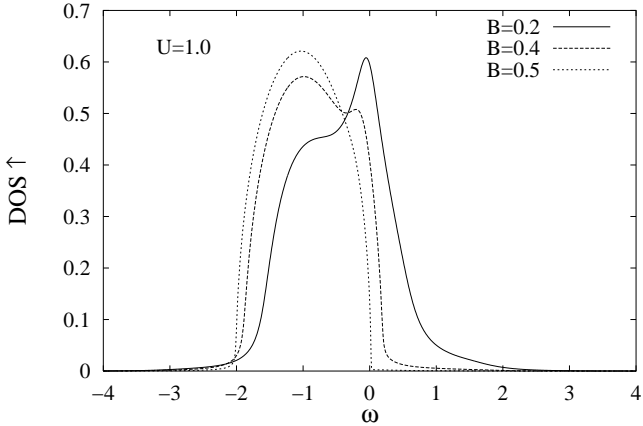,width=6cm, angle=-90}
\caption{\label{fig:dos_weak} DOS of $\uparrow$-electrons at weak coupling
  for different applied fields.} 
\end{figure}

The one-particle instability is realized at weak coupling for $U<U_{c_1}$.
The magnetization decreases and the DOS at the Fermi energy increases
continuously upon decreasing the applied field $B$, Fig.~\ref{fig:dos_weak}.

At a critical point $U_{c_1}\approx1.65$ both conditions
(\ref{instability}) are fulfilled simultaneously and for $U>U_{c_1}$ the
two-particle spin-flip instability (\ref{eq:HF-MF-gamma}) at $B_s(U)>1-U/2$
causes the saturated magnet to get locally unstable and to decay to a
magnetically non-saturated solution.

We use the FLEX approximation with all three channels for the quantitative
description of the non-saturated solution at intermediate coupling.  The
self-energy from the most important electron-hole singlet channel has the
explicit analytic representation
\begin{eqnarray}
  \label{eq:SCh-self-energy}
  &&\Sigma^{eh}_\sigma(\omega_+)= U\int_{-\infty}^0\frac{d\omega'}{\pi}
  \left\{G_{-\sigma}(\omega'+\omega_+)\ \mbox{Im}
  \frac{X_{\sigma-\sigma}(\omega'_+)}{1+X_{\sigma-\sigma}(\omega'_+)}
  \right.\nonumber\\  
  &&\left. +\frac{X_{\sigma-\sigma}(\omega'-\omega_+)}{1+X_{\sigma
        -\sigma}(\omega'-\omega_+)}\ \mbox{Im}\ G_{-\sigma}(\omega'_+)
  \right\} \ 
\end{eqnarray}
where $\omega_+=\omega+i0^+$. Analogous formulae hold for the other
channels where only second-order contribution must correctly be subtracted.

\begin{figure}\hspace*{-10pt}
  \epsfig{figure=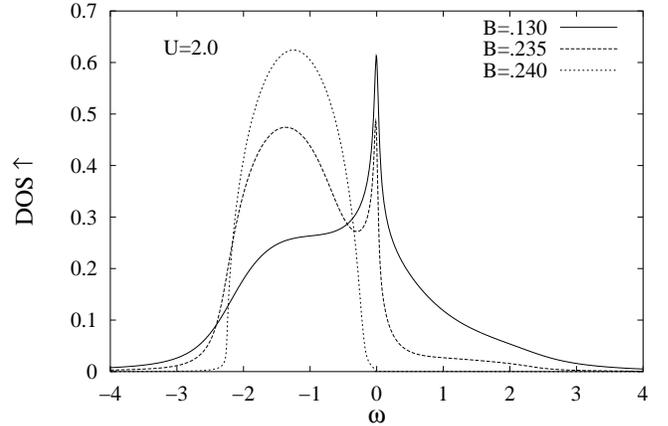,width=6cm, angle=-90}
\caption{\label{fig:dos_inter} DOS of $\uparrow$-electrons at
  intermediate coupling for different applied fields. The local instability
  for the saturated solution lies at $B\approx 0.135$}
\end{figure}

We find that the saturated solution at $B=B_s(U)$ for $U>U_{c_1}$ decays
\textit{discontinuously}, with a jump in the magnetization and energy, to a
metallic solution with pronounced side bands and a sharp maximum in DOS at
the Fermi energy accompanied by a narrow, correlation-induced Kondo
resonance.  Upon increasing the magnetic field, the metallic solution
remains locally stable and preserves the three-peak structure of the
spectral function. The side bands get more pronounced and the central
resonance narrower with a decrease in maximum,
Fig.~\ref{fig:dos_inter}.  The metallic solution gets locally unstable at
an upper critical field $B_u(U)$ at which it again \textit{discontinuously}
goes over to the saturated magnetic insulator. The upper critical field is
determined by a one-sided ($B\searrow B_u$) divergence in the longitudinal
(local) magnetic susceptibility determined by the triplet electron-hole
vertex ${\cal K}$. The central Kondo resonance develops from a maximum to a
sharp edge for the side bands with a jump in the Fermi-level occupation at
the transition point.

From the explicit analytic form of the self-energy contribution
(\ref{eq:SCh-self-energy}) one immediately sees that infinite order
summations are necessary to obtain an instability of the Hartree, fully
spin-polarized solution for $B+U/2>1$. In the fully polarized solution both
$G_{\downarrow}(\omega_+)$ and $X_{\uparrow\downarrow}(\omega_+)$ are real
for $\omega<0$, since the $\downarrow$-DOS is unoccupied. However, a pole
in the vertex function $\Gamma_{\uparrow\downarrow}(\omega_+)$, i.e. when
the condition $1+X_{\sigma-\sigma}(\omega_+)\le0$ is fulfilled for
$\omega<0$, generates a delta-function-like peak near the Fermi energy in
the integrand. A nonvanishing contribution to $\Sigma_{\sigma}$ (i.e. the
self-energy beyond Hartree) emerges even when starting the iterations from
the polarized state. Clearly this would not occur in any weak-coupling
treatment (e.~g. second-order perturbation theory in $U$). Self-consistence
is then needed to establish a locally stable solution.

The coexistence region with two locally stable solutions, metallic and
insulating, emerges due to a global instability of the magnetically
saturated state. We find that instability condition
(\ref{eq:HF-MF-gamma}) is first fulfilled for a magnetization $m$ not
obeying (\ref{eq:HF-MF-m}). The local two-particle instability at
equilibrium is preceded by an instability for a non-equilibrium
magnetization. A first instability appears for $m=0$, but for the
\textit{triplet} channel with $X_{\uparrow\uparrow}$. For the semi-elliptic
DOS we then have an equation $\pi/4U=\frac 23(1-B^2)^{3/2}$ defining an
upper bound, $B_{0}(U)$, for a global instability of the Hartree solution.
The fully saturated magnetic state is hence \textit{globally} stable only
for $B>B_0(U)$ and $B>B_s(U)$.
\begin{figure}\hspace*{-10pt}
  \epsfig{figure=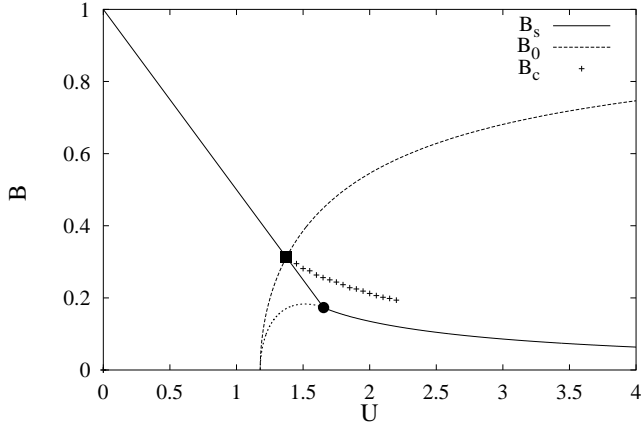,width=6cm, angle=-90}
\caption{\label{fig:coexistence} Local ($B_s$) and global ($B_0$)
  instabilities of the magnetically saturated solution with two
  critical points $U_{c_0}$ (\Pisymbol{pzd}{110}) and $U_{c_1}$
  (\Pisymbol{pzd}{108}). The dashed lines below the critical points show
  the respective instabilities of the non-saturated Hartree solution.}
\end{figure}
However, the actual insulator-to-metal transition does not occur at the
instability of either the insulating or the metallic solution. It is the
energy that decides which solution is at global minimum. The line of the
\textit{first-order} metal-insulator transition, $B_c(u)$, depends on the
approximation and lies in the coexistence region between the boundaries
defined by local instabilities of the insulating and metallic solutions
$B_s<B_c<B_u\le B_0$, respectively\cite{note1}. The line of first-order
transitions ends up at a critical point $U_{c_0}$ where the line of the
global, $B_0(U)$, and of the local, $B_s(U)$, instability meet,
Fig.~\ref{fig:coexistence}. 

To conclude, we have investigated correlated electrons at zero temperature
subjected to an external magnetic field and found that local instabilities of
the insulating, fully saturated magnetic solution at intermediate and
strong coupling are preceded by a \textit{global} instability.  The
saturated magnetic state decays via a \textit{first-order} transition with
a jump in magnetization to a non-saturated, metallic solution. We
explicitly demonstrated the coexistence of metallic and insulating
solutions on a mean-field ($d=\infty$) model, but the same holds also for
finite-dimensional systems near magnetic saturation.  We have to take into
consideration global fluctuations when investigating metal-insulator
transitions and the stability and decay of fully polarized magnets.  We
further showed that the existence and shape of the metallic solution in the
coexistence region is conditioned by the existence of poles in
\textit{two-particle} electron-hole vertices.  To understand the
metal-to-insulator transition behavior in systems with correlated electrons
completely requires to match the magnetic case studied here with the spin
symmetric situation and the Mott-Hubbard transition there.

The work was in part supported by grant No. 202/98/1290 of the Grant Agency
of the Czech Republic (VJ). VJ thanks University of Bremen for hospitality
enabling completion of the present research.




\begin{references}
\bibitem{note0} The atomic limit is contained in IPT only in the
  spin-symmetric case. There are self-energy contributions beyond $U^2$ for
  $n_{\downarrow} \neq n_{\uparrow}$.
\bibitem{note1} We found numerically within FLEX that $B_u<B_0$ for
  $U>U_{c_0}$, but we were unable to exclude the equality for the exact
  theory.
\end{references}
\end{document}